\documentclass[sigplan,screen,authorversion]{acmart}

\usepackage[utf8]{inputenc}
\usepackage[T1]{fontenc}
\usepackage{microtype}
\usepackage[
	disable
]{todonotes}
\usepackage[capitalise,nameinlink]{cleveref}
\usepackage{csquotes}
\usepackage{hyperref}

\usepackage{pgfplots}
\usetikzlibrary{spy}
\usepackage{booktabs}
\usepackage{diagbox}
\usepackage{multirow}
\usepackage{flushend}
\usepackage{float}
\usepackage{balance}

\AtBeginDocument{%
  \providecommand\BibTeX{{%
    \normalfont B\kern-0.5em{\scshape i\kern-0.25em b}\kern-0.8em\TeX}}}

\setcopyright{licensedothergov}
\acmPrice{15.00}
\acmDOI{10.1145/3426425.3426931}
\acmYear{2020}
\copyrightyear{2020}
\acmSubmissionID{sle20main-p15-p}
\acmISBN{978-1-4503-8176-5/20/11}
\acmConference[SLE '20]{Proceedings of the 13th ACM SIGPLAN International Conference on Software Language Engineering}{November 16--17, 2020}{Virtual, USA}
\acmBooktitle{Proceedings of the 13th ACM SIGPLAN International Conference on Software Language Engineering (SLE '20), November 16--17, 2020, Virtual, USA}

\begin{document}

\title[A Precedence-Driven Approach for Concurrent Model Synchronization Scenarios]{A Precedence-Driven Approach for Concurrent Model Synchronization Scenarios using \\ Triple Graph Grammars}


\author{Lars Fritsche}
\orcid{0000-0003-4996-4639}
\affiliation{%
	\institution{Technical University Darmstadt}
	\city{Darmstadt}
	\country{Germany}
}
\email{lars.fritsche@es.tu-darmstadt.de}

\author{Jens Kosiol}
\orcid{0000-0003-4733-2777}
\affiliation{%
	\institution{Philipps-Universität Marburg}
	\city{Marburg}
	\country{Germany}
}
\email{kosiolje@mathematik.uni-marburg.de}

\author{Adrian Möller}
\affiliation{%
	\institution{Technical University Darmstadt}
	\city{Darmstadt}
	\country{Germany}
}
\email{adrian.moeller@stud.tu-darmstadt.de}

\author{Andy Schürr}
\orcid{0000-0001-8100-1109}
\affiliation{%
	\institution{Technical University Darmstadt}
	\city{Darmstadt}
	\country{Germany}
}
\email{andy.schuerr@es.tu-darmstadt.de}

\author{Gabriele Taentzer}
\orcid{0000-0002-3975-5238}
\affiliation{%
	\institution{Philipps-Universität Marburg}
	\city{Marburg}
	\country{Germany}
}
\email{taentzer@mathematik.uni-marburg.de}


\begin{abstract}
Concurrent model synchronization is the task of restoring consistency between two correlated models after they have been changed concurrently and independently.
To determine whether such concurrent model changes conflict with each other and to resolve these conflicts taking domain- or user-specific preferences into account is highly challenging.
In this paper, we present a framework for concurrent model synchronization algorithms based on Triple Graph Grammars (TGGs). 
TGGs specify  the consistency of correlated models using grammar rules; these rules can be used to derive different consistency restoration operations.
Using TGGs, we infer a causal dependency relation for model elements that enables us to detect conflicts non-invasively.
Different kinds of conflicts are detected first and resolved by the subsequent conflict resolution process. 
Users configure the overall synchronization process by orchestrating the application of consistency restoration fragments according to several conflict resolution strategies to achieve individual synchronization goals.  
As proof of concept, we have implemented this framework in the model transformation tool eMoflon.
Our initial evaluation shows that the runtime of our presented approach scales with the size of model changes and conflicts, rather than model size.
\end{abstract}


\newcommand{\marked}{$\text{\rlap{$\checkmark$}}\square$}
\newcommand{\marking}{$\square \rightarrow $ \marked}
\newcommand{\Cdn}{C\textsubscript{dn}}
\newcommand{\Cnd}{C\textsubscript{nd}}
\newcommand{\Cattr}{C\textsubscript{attr}}
\newcommand{\Cnn}{C\textsubscript{nn}}
\newcommand{\Cidn}{C\textsubscript{idn}}
\newcommand{\Cnid}{C\textsubscript{nid}}


\maketitle

\section{Introduction}
\label{chapter:introduction}
Model-driven engineering~\cite{BCW17} has proven to be an effective means to tackle the challenges that accompany the development of modern software systems, which are getting increasingly complex and distributed in nature.
Often more than one model is needed to describe the developed software system from different but overlapping perspectives. 
Keeping these models and various types of traceability relationships between them in a consistent state is a challenging task, often called \emph{model synchronization}.

Model synchronization becomes especially challenging when multiple correlated models are changed concurrently.
In such cases, not all changes can always be propagated between models as some may contradict each other and thus, are in conflict.
This is the case, for example, when a change in one model leads to the deletion of elements in the other whose existence is the prerequisite for changes performed in that second model by another user. 
Yet, even for model changes that are not in conflict, there may be multiple ways to propagate them between models. 

\begin{figure}[h]
	\centering
	\includegraphics[width=\linewidth]{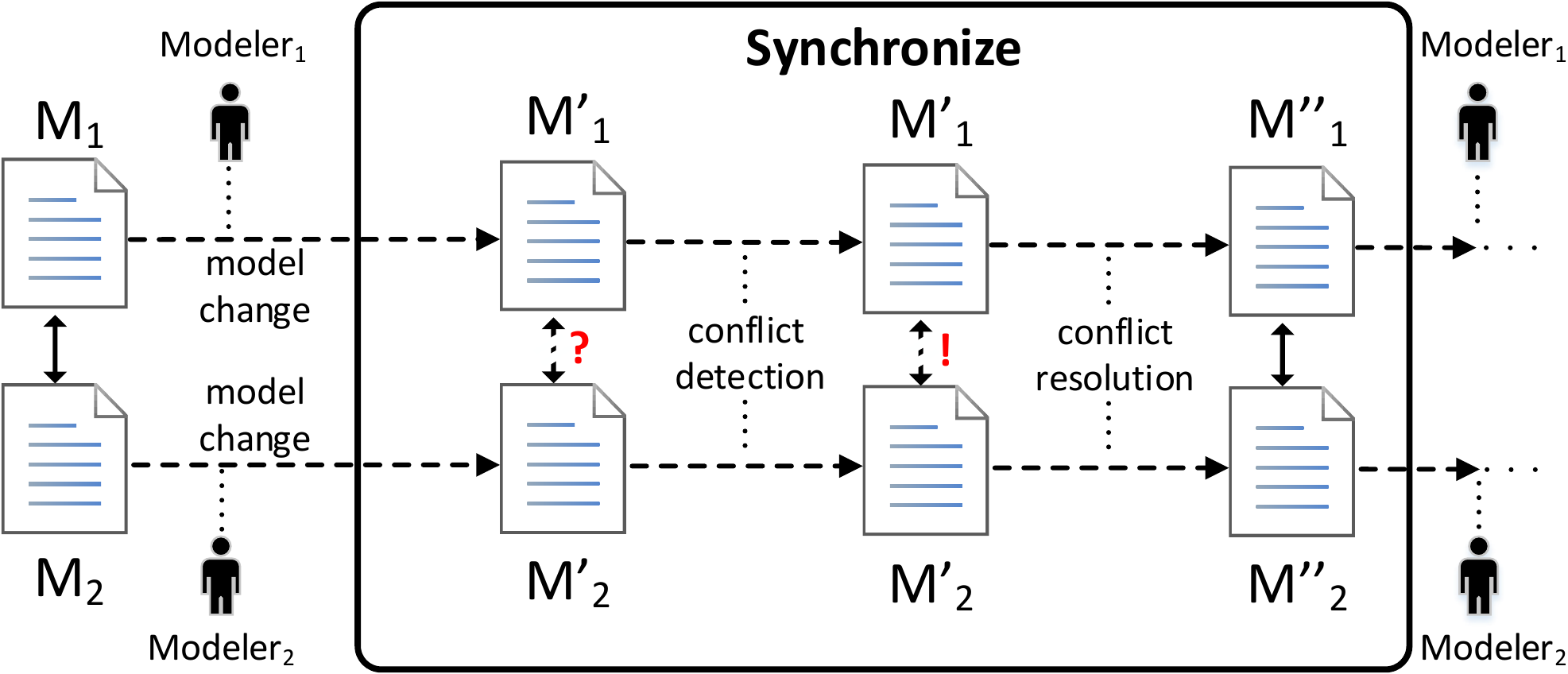}
	\caption{Synchronization Process}
	\label{fig:abstract_process}
\end{figure}

For a modern concurrent synchronization approach, it is of paramount importance to identify synchronization conflicts reliably and give modelers the ability to orchestrate model synchronization processes for guiding the process in accordance with their goals. 
\Cref{fig:abstract_process} gives an overview of a concurrent model synchronization process. 
Consistent interrelated models M$_1$  and M$_2$ are given and changed concurrently. 
The synchronization process identifies all conflicts between these changes and runs a conflict resolution process. 
The expected synchronization result is a consistent pair of models $M^{\prime\prime}_{1}$ and $M^{\prime\prime}_{2}$ that contains all conflict-free changes.

Some approaches provide solely one hard-wired solution of a conflict-detection and -resolution strategy (from a universe of many different options), others come without any formal guarantees for their synchronization results or have an exponential runtime behavior w.r.t. the size of the processed models (cf. \cref{chapter:related_work}).

The contribution of this paper is a framework that simplifies the implementation (orchestration of a family) of concurrent synchronization algorithms with the following properties:
\begin{itemize}
	\item These algorithms are derived from a declarative rule-based formal specification of a model consistency relation in the form of so-called Triple Graph Grammars (TGGs).
	
	\item They come with a number of predefined but extensible conflict-detection and -resolution as well as consistency restoration strategies.
	
	\item Formal properties can be shown using state-of-the-art category-theory- and graph-transformation-based proof techniques (e.g.,~\cite{EEGH15,HEOCDXGE15,Orejas20,FKST20,KFST20}).
	
	\item The intended scope of the effects and potential conflicts of model changes are identified using a TGG-based causal dependency relation for model elements.
	
	\item Scaleability with the size of processed model changes is achieved by limiting the effects of model updates to causally dependent areas in the regarded models and relying on incremental graph pattern matching techniques.
\end{itemize}

In \cref{chapter:related_work}, we give an overview of state-of-the-art concurrent model synchronization approaches.
\Cref{chapter:preliminaries} recalls various concepts related to TGGs. 
\Cref{chapter:conflict_detection,chapter:consistency_restoration} present our concurrent synchronization framework.
The former explains in detail how conflicts are detected, while the latter presents strategies to resolve them and restore consistency.
In \cref{chapter:implementation}, we briefly introduce our implementation. 
Based on this, we evaluate our approach w.r.t. scalability in \cref{chapter:evaluation}. 
\Cref{chapter:conclusion} sums up our contribution and discusses future work. 
In appendices, we give an overview about the situations that can lead to different kinds of conflicts (\cref{{chapter:overview-conflicts-actions}}) and extend an example from \cref{chapter:consistency_restoration} illustrating a possible restoration process (\cref{chapter:conflict-res-ext}). 

\section{Related Work}
\label{chapter:related_work}
In this section, we discuss the state-of-the-art in the field of concurrent model synchronization.
Although, we do not claim completeness of this survey, we are not aware of further works that differ fundamentally from the ones discussed here.
The majority of  the approaches in this field can be categorized into {\em propagation-based}, {\em constraint-based}, and \emph{search-based} approaches.
The approaches considered support \emph{state-based} and \emph{delta-based} model changes; however, they cannot be clearly clustered as some approaches abstract from the way how model changes are described and, therefore, support both state- and delta-based definitions of model changes.

\emph{State-based} approaches~\cite{Xiong08, Xiong09, Xiong11} hold copies of all models to calculate differences, which is not only memory consuming but also scales with the size of the involved models.
In contrast, \emph{delta-based} approaches~\cite{Giese09, Hermann12, Gottmann13, Trollmann17} operate on model changes, which may, e.g., be detected by an incremental pattern matcher. 
In general, delta-based approaches tend to scale better in scenarios with frequent changes but require more bookkeeping, which may have negative effects on memory consumption.

{\em Propagation-based} approaches to concurrent model synchronization use sequential synchronization steps to propagate the changes from one model to the other one followed by a propagation step in the opposite direction.
All propagation-based approaches have severe drawbacks: 
Buchmann et al. \cite{Buchmann2016} employ purely hand-crafted solutions, which do not guarantee any  correctness. 
Especially, they do not show that the synchronization result is still in the given modeling language. 
Some approaches such as \cite{Buchmann2016, Egyed07, Xiong11, Kolovos08, Pierce03, Tratt08, Xiong08, Xiong09} do not consider conflicts between changes on both sides and/or do not provide the means to identify and solve them, which can lead to problems when model changes are overwritten by propagation without asking the modeler.
Some propagation-based approaches are able to tackle the problem of detecting conflicts between parallel updates such as \cite{Trollmann17, Giese09, Xiong11, Hettel08, Hermann12, Gottmann13, Kolovos08} by analyzing if a propagation step contradicts with a model change; in particular, conflict detection happens on-the-fly.
However, as shown by Orejas et al.~\cite{Orejas13}, this propagation-based conflict detection is not deterministic in general. 
Certain conflicts may or may not be detected, depending on the propagation order. 
Other approaches are limited to a confluent set of grammar rules~\cite{Hermann12} or are limited to specific kinds of models such as tree-like hierarchies~\cite{Pierce03}.
Moreover, most of these approaches were either never implemented in a tool or are not any longer available and stable.

\emph{Constraint-based} approaches are often based on a relational specification that can be enforced using tools such as a SAT solver. 
They typically solve problems globally; this means that all possible synchronization solutions are encoded in a search space.
The approach proposed by Macedo et al. \cite{MC16}, for example, finds the closest model that is consistent again. 
Closeness can either be defined w.r.t. graph edit distance or be based on user-defined distance metrics to support user-preferences.
However, this flexibility comes at the price of scalability as constraint-based approaches can often cope with rather small models only.

\emph{Search-based} approaches explicitly explore and find a rich set of synchronization solutions, which can become very expensive with increasing search space.
The work of Cicchetti et al.~\cite{Cicchetti11} aligns itself with that of Macedo et al. in that they calculate all closest sub-models that still conform to a given relational consistency specification.
Focussing on sub-models, there is a potentially large amount of information loss as their approach does not truly incorporate all kinds of model changes.
Orejas et al. \cite{Orejas20} propose a TGG-based approach, where a set of consistency-describing grammar rules is used to find all possible parse trees of the given inter-related models and enriching them with annotations for, e.g., mandatory, removed, added, and no longer covered elements.
These annotations are used to find conflicts, which are resolved using back-tracking to calculate all possible synchronization solutions and to present them to the modeler to choose from.
However, finding all possible synchronization solutions is very expensive and the amount of presented alternative solutions to the user might be overwhelming.
Furthermore, the approach has not been implemented.

In summary, all the approaches discussed have one or more limitations.
We are looking for a concurrent synchronization approach that (1) does not come with severe restrictions concerning the structure of the processed models or the definitions of the regarded consistency relation, (2) finds all kinds of conflicts between concurrent model changes in a deterministic way, (3) allows the modelers to interact with the synchronization process, (4) reliably returns a synchronization result that belongs to the given modeling language, and (5) scales with the size of model changes and conflicts rather than with the model size. 

In this paper, we will present an approach that has all these properties.
However, its scalability comes with the price of the restriction that consistency restoring operations modify only model parts that are causally dependent on those parts that are directly changed by modelers.
Finally, and in contrast to most existing works, we implemented our approach in a state-of-the-art graph transformation tool, namely eMoflon~\cite{Weidmann19}.

\section{Triple Graph Grammars}
\label{chapter:preliminaries}
In this section, we recall \emph{triple graph grammars} (TGGs)~\cite{Schuerr95}, a declarative and rule-based approach to specify the consistency between two modeling languages. 
Being based on (typed attributed) graphs and their transformations as underlying formalism, TGGs are expressive and allow for the development of synchronization solutions with strong formal guarantees~\cite{EEGH15,HEOCDXGE15,Orejas20,FKST20}. 
Moreover, (many of) the operations needed during model synchronization algorithms can automatically be derived from the rules of a given TGG. 
Still, they can be implemented in a scalable way~\cite{ADJKLW17,ABWDKEHSZ20}. 
As (typed attributed) graphs provide a suitable basis to formalize models and their transformations~\cite{BET12}, TGG-based synchronization approaches are directly applicable to models. 
We thus use the terms \enquote{graph} and \enquote{model} interchangeably. 
In the following, the TGG concepts are recalled informally; a formal introduction can be found in, e.g.,~\cite{EEPT06,EEGH15}. 
An informal introduction to graph transformation~\cite{HT20} (including a chapter on model translation and synchronization) appeared recently. 
We illustrate TGGs and the basic ingredients for our synchronization approach using a running example. 

\paragraph{Triple graphs and rules} 
A \emph{triple graph} consists of three graphs: a \emph{source graph}, a \emph{target graph}, and a \emph{correspondence graph} in between that connects source and target graphs via two graph homomorphisms. 
The correspondence graph serves to establish traceability links between correlated elements from source and target graphs. 
In practical applications, the underlying graphs are usually typed and attributed. 
During synchronization processes, the occurring objects may become {\em partial triple graphs}~\cite{FKST19,KFST20}: 
A user may have deleted an element that was referenced by a correspondence morphism. 
Partial triple graphs still consist of three graphs; the graph homomorphisms connecting the correspondence graph with the source and target graph, however, may be partial, i.e., contain dangling references.

\begin{figure}[h]
	\centering
	\includegraphics[width=\linewidth]{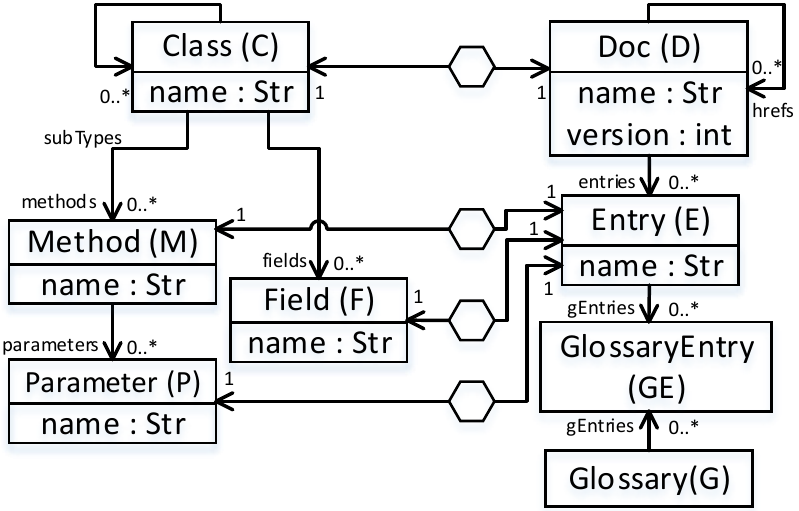}
	\caption{Metamodel of the running example}
	\label{fig:metamodel}
\end{figure}

As a running example for a TGG, we define the consistency between a Java abstract syntax tree (AST) model (source) and a documentation model (target) as depicted in \cref{fig:metamodel}. 
This figure shows a metamodel (represented as a triple type graph) 
that declares the general syntax of 
models. 
The Java AST model consists of \textsf{(Sub-)Classes} containing \textsf{Methods} with \textsf{Parameters} and \textsf{Fields}, while the documentation model consists of \textsf{Doc(ument)s} with \textsf{hyper references (href)} to other \textsf{Docs}.
Furthermore, a \textsf{Doc} contains \textsf{Entries} referencing \textsf{Glossary Entries}, that again are contained in a \textsf{Glossary}. 
Note that some elements have a \textsf{name} attribute, while \textsf{Docs} additionally store their \textsf{version} number.
The correspondence types are depicted as hexagons referencing types of both, the Java and the documentation model, pair-wisely.

A \emph{triple rule} consists of two triple graphs $L$ and $R$ (typed over the given triple type graph), called \emph{left-hand side} (LHS) and \emph{right-hand side} (RHS), respectively, such that their intersection $K = L \cap R$ is a triple graph again. 
Intuitively, the difference between $L$ and $K$ specifies all the elements to be deleted by an application of the rule, and the difference between $R$ and $K$ specifies all the elements to be created. 
Additionally, rules may be equipped with \emph{negative application conditions} (NACs) that specify forbidden context in whose presence the rule is not applicable. 
Finally, we allow rules to be equipped with constraints concerning the attribute values. 
In this paper, we restrict them to be equations involving attribute values of corresponding elements only. 

A \emph{triple graph grammar} consists of a start graph (that is usually the empty graph) and a set of non-deleting rules, i.e., rules, where the LHS is a sub-triple graph of the RHS. 
The \emph{language} defined by a TGG consists of all graphs derivable by an application sequence using its rules beginning at its start graph. 
Hence, a pair of models is consistent w.r.t. a given TGG if and only if a correspondence graph exists that extends the two models to a triple graph in the TGG\rq{}s language.
Rewriting of partial triple graphs can be introduced analogously~\cite{KFST20} and will be used to capture the semantics of synchronization operations manipulating dangling references.

\begin{figure}[h]
	\centering
	\includegraphics[width=0.85\linewidth]{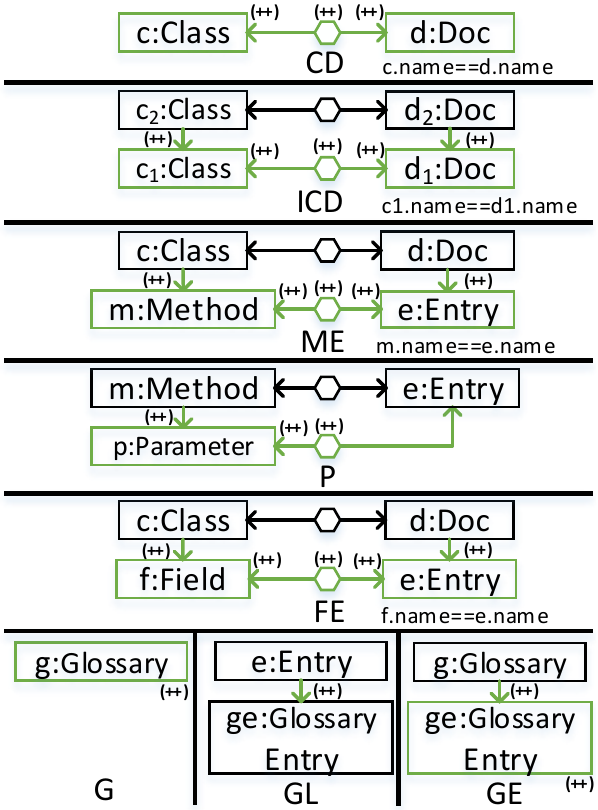}
	\caption{TGG Rules}
	\label{fig:ruleset}
\end{figure}

\begin{figure}[h]
	\centering
	\includegraphics[width=0.55\linewidth]{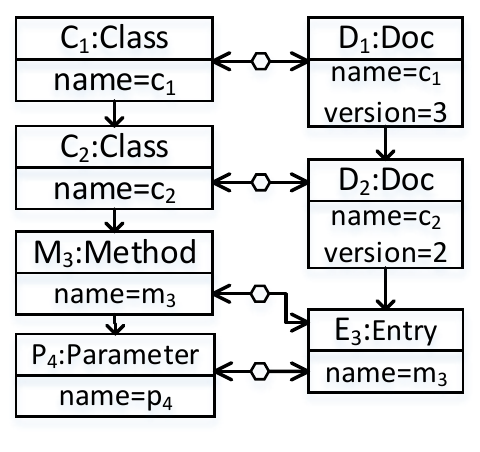}
	\caption{Exemplary Model}
	\label{fig:simple_corr}
\end{figure}

\Cref{fig:ruleset} depicts the rule set of our running example consisting of 8 TGG rules. 
They are displayed in an integrated fashion, i.e., as a single graph. 
The black, unmarked elements constitute the LHS of the rule, i.e., the context that has to exist for a rule to be applicable. 
Green elements annotated with (++) are to be created when the rule is applied.
The rule \emph{CD} has no precondition and thus may be used to generate \textsf{Classes} with corresponding \textsf{Docs} arbitrarily often. 
The rule ICD creates a \textsf{Sub-Class} and a corresponding \textsf{Doc} with a \textsf{hyper-reference} when a \textsf{Class} with a corresponding \textsf{Doc} already exists.
Note that we create the subClass link together with the sub-class, which implicitly forbids multiple inheritance.
The rules \emph{ME} and \emph{FE} create \textsf{Methods}, resp. \textsf{Fields}, with corresponding \textsf{Entries}. 
Rule \emph{P} creates a \textsf{Parameter} that corresponds to an already existing \textsf{Entry}.
Finally, the rules \emph{G}, \emph{GE}, and \emph{GL} create a \textsf{Glossary} together with \textsf{Glossary Entries} and links from \textsf{Entries} to \textsf{Glossary Entries}.
These rules only act on the documentation model (the target side); the created elements do not have corresponding Java elements. 
Several rules (\emph{CD}, \emph{ICD}, \emph{ME}, and \emph{FE}) are equipped with attribute conditions. 
In each case, the condition declares that the names of the newly created elements should be equal. 
\Cref{fig:simple_corr} depicts a simple model that can be created applying first rule \emph{CD} followed by applications of rules \emph{ICD}, \emph{ME}, and \emph{P}.

\begin{figure}[h]
	\centering
	\includegraphics[width=0.7\linewidth]{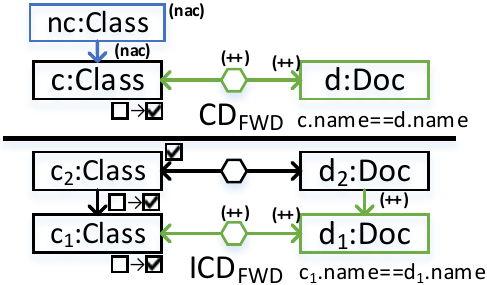}
	\caption{Exemplary TGG Forward Rules}
	\label{fig:forward_rules}
\end{figure}

\paragraph{Rules derived from TGG rules}
Triple rules 
can be used to create consistent models from scratch. 
For scenarios like \emph{model translation} and \emph{model synchronization}, suitable kinds of rules can be derived from the given TGG. 
First, a TGG rule can be \emph{operationalized} to support forward (source $\rightarrow$ target) and backward (target $\rightarrow$ source) translations.
\Cref{fig:forward_rules} depicts the forward operationalized rules \emph{CD}\textsubscript{FWD} and \emph{ICD}\textsubscript{FWD}, which are created by converting all green source elements to context as we consider them to already exist.
To prevent elements from being translated twice, we introduce annotations: \marking{} indicates that this element is still untranslated and  applying the rule marks this element as translated.
Consequently, \marked{} indicates that the annotated element must be translated before applying this rule.
Note that \emph{CD}\textsubscript{FWD} contains a NAC, depicted in blue and annotated with (nac), that forbids the \textsf{Class} \emph{c} to be translated into a corresponding \textsf{Doc} when it is a \textsf{Sub-Class} of another \textsf{Class}. 
This kind of NAC is also called a \emph{filter NAC}~\cite{Hermann10,KLKS10} and adjusts the translation process to avoid dead-ends.  
If a \textsf{Sub-Class} is translated using \emph{CD}\textsubscript{FWD} (without that NAC), for example, there is not any other forward rule that translates the remaining link to its \textsf{Super-Class}.
The creation of all other forward and backward rules can be done analogously. 

\begin{figure}[h]
	\centering
	\includegraphics[width=0.55\linewidth]{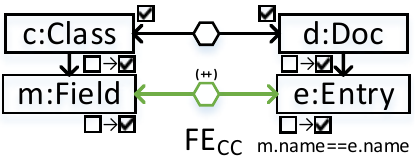}
	\caption{Exemplary TGG Consistency Check Rule}
	\label{fig:cc_rule}
\end{figure}

Another useful operationalization of TGG rules is referred to as {\em consistency check rules}. 
They detect whether yet untranslated elements in a source and a target model can be considered as correlated. 
If so, they create correspondence links. 
While these kinds of rules have been used to compute (maximal) correspondence relations between previously unrelated models from the source and target domain~\cite{Erhan2018}, we employ them only \emph{locally} to detect whether independently added elements on source and target side can be considered as corresponding to each other. 
We refer to this process as \emph{local-CC}.
\Cref{fig:cc_rule} depicts an exemplary consistency check rule that is derived from rule \emph{FE}. 
Moreover, we project these consistency check rules to their source and target parts only to obtain \emph{source} and \emph{target patterns}, which we will use for conflict detection purposes later on. 

Finally, there are \emph{short-cut rules} that are synthesized from TGG rules by means of a special kind of sequential rule composition operator~\cite{FKST18}. 
Applying a short-cut rule replaces one TGG rule application by another one while allowing to preserve selected elements (instead of deleting and recreating them).
Short-cut rules allow for advanced editing of models while preserving information in the process. 
Forward or backward operationalizing a short-cut rule results in \emph{repair rules} that allow to directly propagate the edit the short-cut rule specifies on the source side to the target (and conversely). 
The crucial point is that short-cut rules specify complex (language-preserving) edits one cannot immediately perform using the original rules of the given TGG. 
Hence, their derived repair rules are often suitable to directly propagate \enquote{free} user edits. 
Short-cut rules do not need to be non-deleting and their derived repair rules may act on partial triple graphs. 
For details on their construction, conditions on language-preserving applications, and application in unidirectional model synchronization, we refer to~\cite{FKST18,FKST19}.

\begin{figure}[h]
	\centering
	\includegraphics[width=0.7\linewidth]{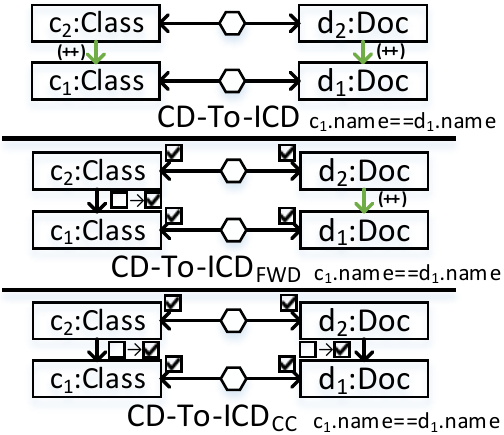}
	\caption{(Operationalized) Short-Cut Rule}
	\label{fig:short_cut_rule}
\end{figure}

An example of a short-cut rule is given in \cref{fig:short_cut_rule}, where the short-cut rule \emph{CD-To-ICD} transforms an application of \emph{CD} to one of \emph{ICD}.
\emph{CD-To-ICD\textsubscript{FWD}} shows the forward operationalized short-cut rule, which directly propagates an edge that was newly inserted between two classes. 
Furthermore, also short-cut rules can be operationalized to obtain consistency check operations such as \emph{CD-To-ICD$_{CC}$}.

\paragraph{Precedence graph}
Given a consistent triple as depicted in \cref{fig:simple_corr}, we can infer a \emph{precedence graph} (PG) that describes with which TGG rule applications this triple can be derived and how those depend on one another. 
This means that a PG describes the aforementioned causal dependency relationship for model elements.
Formally, a precedence graph is based on so-called \emph{consistency patterns}. 
A consistency pattern is just the RHS of one of the rules of the given TGG. 
Applying a rule to a triple graph, one obtains a homomorphism from the RHS of the rule to the resulting triple graph: 
This homomorphism maps the elements of the LHS of the rule to the elements they have been matched to for applying the rule. 
The elements of its RHS that do not belong to the LHS (i.e., the elements to be created) are mapped to the newly created elements. 
Thus, given a sequence of rule applications, 
one obtains a family of homomorphisms from the rules\rq{} RHSs to the triple graph. 
These homomorphisms are in a natural \emph{dependency relation}: 
One homomorphism is dependent on another one if the former matches an element that the underlying rule application of the latter one creates. 
Moreover, these homomorphisms \emph{cover} the triple graph in the following sense: 
Every element of it is matched exactly once by a rule element to be created, i.e., by an element of RHS\,$\setminus$\,LHS of one of the rules. 
We will also say that this match \emph{explains} or \emph{accounts for} the element. 
Conversely, given a TGG and some triple graph, if there is a family of homomorphism from the consistency patterns to this triple graph such that the dependency relation is acyclic and the family covers the triple graph in the above sense, the triple graph belongs to the language of the TGG (for a formalization and proof, we refer to~\cite[Lemma~4]{Erhan2018}). 
This is what we define to be a \emph{precedence graph} (PG) for a triple graph $M$: an acyclic graph where the nodes are homomorphisms from consistency patterns to $M$ such that $M$ is covered by them, and edges are their dependencies.
Note that the dependency relation stored in a PG induces a causal dependency relation on elements of the triple graph: 
An element $x$ (node or edge) depends on an element $y$ if the element $y$ is matched by the rule that creates $x$. 
Similar information (about dependency between and coverage of elements) has been used by Kehrer to lift atomic model changes to the level of \emph{edit scripts}~\cite{Kehrer15}.

\begin{figure}[h]
	\centering
	\includegraphics[width=0.6\linewidth]{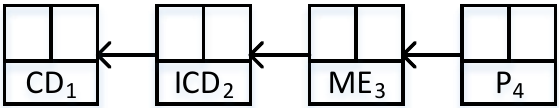}
	\caption{Exemplary Precedence Graph}
	\label{fig:simple_pg}
\end{figure}

\Cref{fig:simple_pg} depicts the PG for the model in \cref{fig:simple_corr} with each node corresponding to a TGG rule application where the name is based on the TGG rule name and an index representing the indices of created elements. 
The boxes inside the nodes represent the state of created elements of the corresponding rule application on source (left box) and on the target (right box) side.
In the next chapter, we will introduce annotations for these boxes that will help us detecting conflicts.

\section{Conflict Detection}
\label{chapter:conflict_detection}
In this section, we present our approach to the detection of conflicts during model synchronization. 
We assume the following general setting (compare Fig.~\ref{fig:abstract_process} again): 
A TGG is fixed; it defines a consistency relation between two modeling languages. 
Our concurrent synchronization process starts with a pair of consistent models $M_1,M_2$, or, somewhat more formally, a triple graph belonging to the language of the given TGG (i.e., $M_1$ is the source and $M_2$ the target graph). 
This pair of consistent models comes with a precedence graph $PG$.  

Both models are changed independently by two modelers, resulting in models $M_1^{\prime}$ (source side) and $M_2^{\prime}$ (target side).
Compared to $M_1$, in $M_1^{\prime}$ some elements may have been added, some deleted, and some attribute values may have changed; the same holds for $M_2$ and $M_2^{\prime}$. 
We call this change a \emph{(model) delta} and speak of \emph{source} or \emph{target delta} when we want to refer to only one of these changes.  
We do not make any assumptions on how they have been performed; we only assume that there is a way to identify the remaining elements of $M_1$ and $M_2$ with their counterparts in $M_1^{\prime}$ and $M_2^{\prime}$, respectively. 

Our goal is to find models $M_1^{\prime\prime}$ and $M_2^{\prime\prime}$ that are consistent, i.e., which are source and target graphs of a triple graph of the given language. 
Moreover, $M_1^{\prime\prime}$ and $M_2^{\prime\prime}$ should not differ too much from $M_1^{\prime}$ and $M_2^{\prime}$, respectively. 
In general, there is no unique solution for this problem, even if requiring the distance between $(M_1^{\prime\prime},M_2^{\prime\prime})$ and $(M_1^{\prime},M_2^{\prime})$ to be \emph{minimal} (according to some metric). 
Therefore, we provide modelers with the possibility to orchestrate the synchronization process, leading to individually defined outcomes. 

To compute a pair of consistent models, we first extend and annotate the precedence graph to obtain a \emph{delta precedence graph} (DPG). 
This graph comprises of information regarding which parts of the PG have been affected by the deltas and how the individual changes can (locally) be propagated to the respective other side. 
{\em Our guiding principles for synchronization are to only change the models where the delta makes this necessary and to preserve as many of the model changes as possible, e.g., to not delete newly added elements and to not recreate deleted ones} (compare, e.g.,~\cite{CGMS17,TOLR17} for these principles). 
Whenever it is not possible to simultaneously preserve user-changes on the source and on the target model such that the model remains in the language of the TGG, we call this a \emph{conflict}. 
Our synchronization process first analyzes the delta precedence graph for conflicts and subsequently propagates the source and target deltas to the respective other side while resolving the detected conflicts according to an orchestration given by the user. 

In this section, we introduce delta precedence graphs and different types of conflicts and illustrate them using our running example. 

\begin{figure*}[th]
  \centering
  \includegraphics[width=\linewidth]{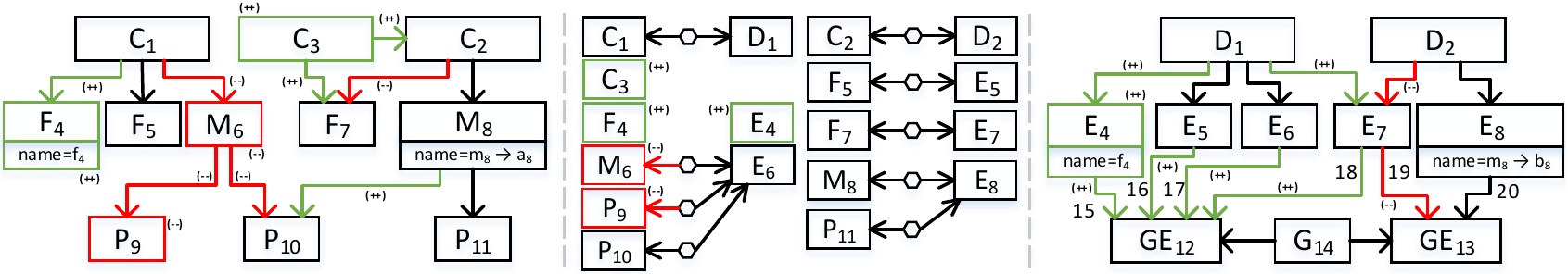}
  \caption{Running Example -- Model and Delta}
  \label{fig:ref-model_delta}
\end{figure*}

\Cref{fig:ref-model_delta} shows an example of a concurrent model change applied to a formerly consistent model graph. 
In the source graph of the original model, there are two \textsf{Classes} (C$_1$ and C$_2$) that have a \textsf{Field} and \textsf{Method} each.
Furthermore, M$_6$ has two \textsf{Parameters} and M$_8$ has one.
The target graph has two \textsf{Docs} (D$_1$ and D$_2$) that contain two \textsf{Entries} each, where the \textsf{Entries} contained in D$_2$ reference the \textsf{Glossary Entry} GE$_{13}$.
GE$_{13}$ and GE$_{12}$ are contained in the \textsf{Glossary} G$_{14}$.
Elements with the same indices on both sides have a correspondence link, except for \textsf{Parameters}, which are connected to those \textsf{Entries} that correspond to their \textsf{Method}.

Several concurrent{\em model changes} have taken place with different impacts and issues.
First, the \textsf{Field} F\textsubscript{7} is pulled up into a newly created \textsf{Super-class} C$_3$ of C$_2$. 
In the target graph, the corresponding \textsf{Entry} E$_7$ is moved to D$_1$, which does not correspond to C$_3$.
Also, a new \textsf{Field} F$_4$ and a new \textsf{Entry} E$_4$ are created within elements C$_1$ and D$_1$, respectively. They have the same names.
Additionally, M$_6$ is deleted together with its \textsf{Parameter} P$_9$, while the \textsf{Parameter} P\textsubscript{10}  was moved to M$_8$.
However, while M$_6$ is deleted in the source graph, the corresponding target element E$_6$ as well as E$_4$, E$_5$, and E$_7$ are linked to GE$_{12}$.
Finally, the names of corresponding model elements M$_8$ and E$_8$ are changed to different values. 
Similarly to our notation of rules, in \cref{fig:ref-model_delta}, newly added elements are depicted in green and marked with $(++)$, deleted ones in red and marked with $(--)$, and attribute changes are indicated via an arrow $\to$. 

\begin{figure*}[h]
	\centering
	\includegraphics[width=.8\linewidth]{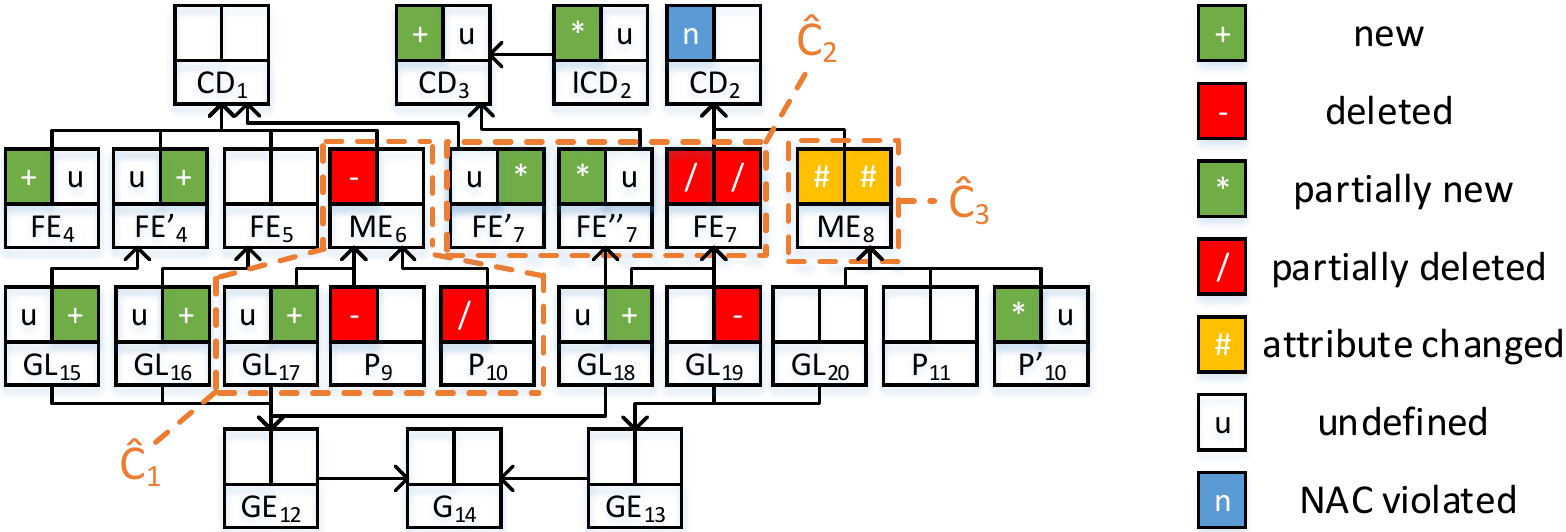}
	\caption{Running Example -- Delta Precedence Graph}
	\label{fig:rex-precedenceGraph}
\end{figure*}

\subsection{Delta Precedence Graphs}
A \emph{delta precedence graph} extends a precedence graph by information on how a delta affected the validity of the precedence graph. 
For elements added in the delta, we need to find suitable TGG-rule applications that could have created these elements; potentially correlating elements have to be created on the other side. 
Moreover, consistency matches of the precedence graph may have been invalidated (or \enquote{broken}) in three different ways: (i) attribute values have been changed such that an attribute condition of a TGG rule is violated, (ii) elements that are covered by a consistency match
have been deleted, and (iii) elements have been added so that already existing elements have to be parsed anew since a NAC is violated.
Elements that are no longer covered by a consistency pattern since the formerly covering one is \enquote{broken} have to be matched anew as well.
All this information is collected in a DPG. 
In the definition of DPG, we call an element of the updated model \emph{unpropagated} if one of the following two cases applies: 
\begin{enumerate}
	\item This element has been newly added by one of the user edits, or
	\item there is a consistency pattern that explains how this element has been created (i.e., the underlying rule application created this element). However, this rule application has been rendered invalid by the user edits (either, because one of its filter NACs is violated now or because a matched element 
has been deleted).
\end{enumerate}

\begin{definition}[Delta precedence graph]
	Let a TGG and a pair of consistent models $(M_1,M_2)$ 
	together with a precedence graph $PG$ for it be given. 
	The \emph{delta precedence graph} (DPG) for $(M_1,M_2)$, $PG$, and a delta (consisting of sequences of graph changes leading from $M_1$ to $M_1^{\prime}$ and $M_2$ to $M_2^{\prime}$, respectively) consists of 
 the nodes of the given precedence graph $PG$ and 
a set of new nodes consisting of matches for source and target patterns that are such that at least one unpropagated element is matched by an element the underlying rule of the pattern creates.
	Edges are, again, defined via dependencies. 

		Moreover, the nodes of the DPG have \emph{source} and \emph{target annotations} over the alphabet $\{+, -, *, /, \#, u, n \}$ according to the following rules:
	\begin{itemize}
		\item Nodes stemming from the original PG, i.e., nodes indicating formerly valid rule applications, are annotated on the source side in the following way:
		\begin{itemize}
			\item annotation \enquote{$-$} whenever all elements the underlying rule application created on the source side have been deleted by the source edit; 
			\item annotation \enquote{$/$} whenever some (but not all) elements the underlying rule application created on the source side have been deleted by the source edit or if the match for this consistency pattern is broken because context elements are missing; 
			\item annotation \enquote{$\#$} whenever some attribute values were changed by the source edit such that at least one attribute constraint of the underlying rule application is violated; 
			\item annotation \enquote{$n$} whenever the source edit added a new element that introduces a violation of a source NAC of the forward rule of the rule from which the consistency pattern is derived. 
		\end{itemize}
		The target annotation is defined completely analogously. 
		A node may also have more than one such annotation, which is why we will refer to sets of annotations in the following.
		However, for simplicity reasons, nodes in our running example only have sets of size 1.
		\item Whenever all nodes that the underlying rule created on the source side are matched to unpropagated elements, a node corresponding to a match for a source pattern is annotated with \enquote{$+$} on the source and with \enquote{$u$} on the target side. 
		Symmetrically, the annotations for such a target pattern are switched. 
		\item Whenever at least one node on the source side, created by the underlying rule, is matched to an already propagated element that was part of another formerly intact consistency match, a node corresponding to a match for a source pattern is annotated via \enquote{$*$} on the source and \enquote{$u$} on the target side.  
		Symmetrically, the annotations for such a target pattern are switched.
	\end{itemize}
	
\end{definition}
\Cref{fig:rex-precedenceGraph} depicts the DPG that corresponds to the model changes described above. 
One of the nodes FE\textsubscript{7}, for example, is marked with \enquote{$/$} on both source and target side because the edges created by the corresponding rule application have been deleted on both sides; however, nodes F\textsubscript{7} and E\textsubscript{7} are preserved. 
The two new nodes indexed with FE$_{7}^{\prime}$ and FE$_{7}^{\prime\prime}$ denoting a new source and target pattern match, respectively, indicate by which rule the now unaccounted nodes F\textsubscript{7} and E\textsubscript{7} could have been created. 
The consistency restoration has to check whether they can be combined into a single TGG-rule application creating the two correlated elements simultaneously. 
Node ME\textsubscript{8} is annotated with \enquote{$\#$} on both sides as the attribute values have been changed on both sides and the constraint requiring equal names is  violated now. 
As a last example, the node CD\textsubscript{2} is annotated with \enquote{$n$} on the source side: 
Due to the newly introduced inheritance edge, it is no longer possible for the class C\textsubscript{2} to be created using rule \emph{CD}. 
Thus, this node becomes unpropagated; the node indexed with ICD\textsubscript{2} indicates a new possibility to parse this node.

\subsection{Conflicts}
As discussed above, the annotations of a DPG indicate some synchonization actions 
 that have to take place; as long as the dependency is respected (i.e., is still acyclic), a consistent triple graph is restored as soon as every annotation has been dealt with. 
Our definition and treatment of conflicts is based on the already mentioned idea of change-preservation. 
Whenever a user edit directly affected an element, this effect should be preserved. 
This means that newly created elements or elements whose attribute values have been changed are \emph{intended to persist}. 
Deleted elements are intended to remain deleted. 
A conflict is a situation where all options available to propagate a certain change require to undo another one.
We classify such conflicts based on the annotations occurring in DPGs and use the term \emph{conflict scope} to refer not only to a conflict but also to other elements that depend on how the conflict is resolved. 
In the following, we assume a pair of models that was originally consistent, a PG for it, a model delta, and the induced DPG to be given. 
In particular, our approach to conflict detection is \emph{static} and, as such, \emph{deterministic} for a fixed PG and delta.
We illustrate all kinds of conflicts using our example in \cref{fig:rex-precedenceGraph} (where the \emph{conflict scopes} $\hat{C}_1$, $\hat{C}_2$, and $\hat{C}_3$ are indicated by dashed orange lines).

\paragraph{Preserve-delete conflict}
A preserve-delete conflict is a situation where one of the deltas deletes a certain element whereas its corresponding element is used in the other delta and thus, intended to persist: 
A \emph{potential preserve-delete conflict} is a node of the DPG where \enquote{$-$} or \enquote{$/$} belongs to the source or target annotation (except for the case where both annotations are \enquote{$-$}).
It is a \emph{preserve-delete conflict} if there is a newly added element on the other side or an element whose attribute value has been changed such that, by propagating the deletion, all patterns that would have been able to create that element vanish. 
Its \emph{scope} is the node itself and all nodes transitively depending on it. 

$\hat{C}_1$ depicts a {\em preserve-delete conflict} that is characterized by a full deletion (\enquote{$-$}) in one precedence node on the source side and a creation of new elements (\enquote{+}) in one of its dependent nodes on the target side.

Here, M$_6$ is deleted on the source side, but a new reference is added between E$_6$ and GE$_{12}$ on the target side. 
The conflict scope $\hat{C}_1$ includes changes that need not directly conflict with each other, but have to be considered when resolving the conflict.
For example, not revoking P$_9$ implies not revoking M$_6$. 

\paragraph{Correspondence preservation conflict}
A  cor\-re\-spond\-ence preservation conflict is a situation where deltas modify corresponding elements on both source and target side such that it is not possible to restore the consistency without either deleting the correspondence relationships between the affected elements (and creating new correspondence relationships to different elements, which in the general case have to be created, too) or without discarding changes in the source or target model. 
In our example, $\hat{C}_2$ is such a conflict where F$_7$ is moved to C$_3$ while E$_7$ is moved to D$_1$. 
However, C$_3$ and D$_1$ do not correspond to each other and thus there is no TGG consistency pattern that can correlate these changes.
Hence, we have to decide whether to revoke the relocation of either F$_7$ or E$_7$.

Every node of the DPG where \enquote{$n$} or \enquote{$/$} belongs to the source and target annotation is a \emph{potential correspondence preservation conflict}. 
It is a \emph{correspondence preservation conflict} if there exists no intact TGG consistency pattern match that again covers and relates these elements. 
Its \emph{scope} is the node itself as well as nodes with a \enquote{$*$} annotation whose corresponding rule applications create some elements that are also created by the rule application of the node itself.

\paragraph{Attribute change conflict}
An attribute change conflict is a special case of a correspondence preservation conflict, where attribute values of until now corresponding source and target elements have been changed in such a way that the attribute values on both sides are no longer consistent. 
In our example, $\hat{C}_3$ is an attribute change conflict that occurs in node ME$_8$. 
The name of M$_8$ is changed from m$_8$ to a$_8$, while that of E$_8$ is changed from m$_8$ to b$_8$.

If both names would have been changed equally, no conflict would have been detected and the incremental pattern matcher would not detect any broken rule application.
Every node of the DPG where \enquote{$\#$} belongs to the source and the target annotation is a \emph{potential attribute change conflict}. 
It is an \emph{attribute change conflict} if furthermore an attribute constraint is violated because both attribute values were changed.

In \cref{chapter:overview-conflicts-actions}, we provide a overview of how the annotations of a node in a DPG relate to potential conflicts.

\section{Consistency Restoration}
\label{chapter:consistency_restoration}
In this section, we present a catalog of concurrent synchronization fragments, which can be orchestrated and executed sequentially to restore the consistency of the model under change in a concurrent synchronization scenario.
This catalog includes the previously introduced operationalizations of TGG rules and three pre-defined conflict resolution strategies. Users can orchestrate their specific consistency restoration processes using these fragments. 
Each fragment processes certain kinds of annotated DPG nodes in order to propagate changes and resolve conflicts, which has a direct effect on the models.

\paragraph{Concurrent synchronization fragments}
In the following, we present all the concurrent synchronization fragments of our catalog, show an example orchestration of them which is quite typical, and apply this orchestration to our running example. 
All the following fragments except for \emph{Resolve Conflict} are applied to elements only that do not belong to a conflict. \emph{Resolve Conflict} is applied to each conflict (scope). 
Furthermore, if a fragment is applied, it processes all feasible PG nodes until no unprocessed one can be found.

\begin{itemize}
	\item \emph{\textbf{Local CC}} is used to find and correlate newly added elements in the source and target graphs that may correspond to each other. This fragment processes pairs of precedence nodes annotated with \textbf{(+|u)} and \textbf{(u|+)} as these belong to newly added and yet unprocessed elements.
	If it is chosen, it has to be applied before \emph{Translate} as this also processes newly added elements that are no longer available afterwards.
	\item \emph{\textbf{Translate}} is used to translate newly added elements from any side to the opposite side and thus complete the rule application by applying a forward or backward operationalized TGG rule. 
	This fragment processes precedence nodes annotated with \textbf{(+|u)} and \textbf{(u|+)} as these belong to newly added and yet unprocessed elements.
	\item \emph{\textbf{Repair}} employs short-cut rules to fix broken precedence nodes. 
	Forward and backward operationalized short-cut rules propagate complex changes from any side to the other. 
	Consistency check operationalized short-cut rules allow to find corresponding complex changes on both sides and resolve them if possible. 
	This fragment  processes precedence nodes that are annotated with \enquote{\textbf{n}} or \enquote{\textbf{/}} on one or both sides and related nodes annotated with \textbf{(*|u)} or \textbf{(u|*)} as these indicate that while the rule application has been violated, some of the remaining elements are to be preserved.
	\item \emph{\textbf{Resolve conflict}} is applied to each conflict (scope) and can be configured for each type of conflict that we identified in the previous chapter. 
	We can call \emph{Translate}, \emph{Repair} and \emph{Propagate}, where
	\emph{Repair} can be used to reduce the conflict size beforehand, while \emph{Translate} and \emph{Propagate} can only be used after the conflict (scope) has been resolved because both include propagation steps that can only be executed after the conflicts cause has been resolved.
	For resolving a conflict (scope), we offer three pre-defined strategies:
	\begin{itemize}
		\item \emph{\textbf{Take Source}} discards all the changes on the target side.
		\item \emph{\textbf{Take Target}} discards all the changes on the source side.
		\item \emph{\textbf{Preserve}} discards all the deletions that block newly added elements from being propagated.
	\end{itemize}
	We allow modelers to implement conflict evaluation functions that if evaluated true will trigger one of these pre-defined strategies (e.g., if more source elements were deleted than target elements, apply \emph{\textbf{Take Source}}).
	All applied fragments within \emph{Resolve Conflict} are applied to elements only that belong to the current conflict (scope).
	\item \emph{\textbf{Rollback}} revokes rule applications in cases where all the deletions were performed consistently on both sides or only on one side, e.g., all green source elements were deleted.
	In the latter case, the other side remains untouched but has to be deleted as a consequence. 
	This fragments processes nodes annotated solely with \enquote{\textbf{-}} on one or both sides as this indicates a consistent deletion of all green elements.
	\item \emph{\textbf{Propagate}} applies \emph{Repair} first to fix broken matches rather than revoking them. 
	Then, it applies \emph{Rollback} to revoke rule applications that have been consistently deleted on one side.
	Finally, \emph{Translate} is applied and translates newly added elements to the opposite side.
	From our experience, calling these fragments in that specific order is a good choice for an separate fragment as it yields a sequential synchronization control flow.
	\item \emph{\textbf{Clean up}} deletes all elements from source, correspondence, and target graphs that are currently inconsistent w.r.t. our TGG. 
	This fragment can only be called at the end of a synchronization process but can also be omitted if the user decides to eliminate inconsistencies later on or generally wants to tolerate some inconsistencies. 
	Note that through this fragment we can guarantee correctness of our results (i.e., output belonging to the language of the given TGG) but no kind of optimality of them.
\end{itemize}
To restore the consistency of the model in our running example, we choose the following orchestration of fragments to be applied in the given sequential order: \emph{Local CC} $\rightarrow$ \emph{Translate} $\rightarrow$ \emph{Repair} $\rightarrow$ \emph{Resolve Conflict \{\emph{Repair} $\rightarrow$ \emph{Take Source} $\rightarrow$ \emph{Propagate}\}} $\rightarrow$ \emph{Propagate} $\rightarrow$ \emph{Clean up}.
For simplicity reasons, we resolve all three conflicts ($\hat{C}_1$,  $\hat{C}_2$,  $\hat{C}_3$ from \cref{fig:rex-precedenceGraph}) uniformly.
However, a modeler can indeed (manually or programmatically) implement his own strategy to choose a strategy for each detected conflict.
Note that no more information has to be given than the order in which these fragments are to be applied and that a modeler does not need to specify which elements are to be handled by a fragment as this is intrinsic for each fragment as specified above.


\begin{figure}[h]
	\centering
	\includegraphics[width=0.8\linewidth]{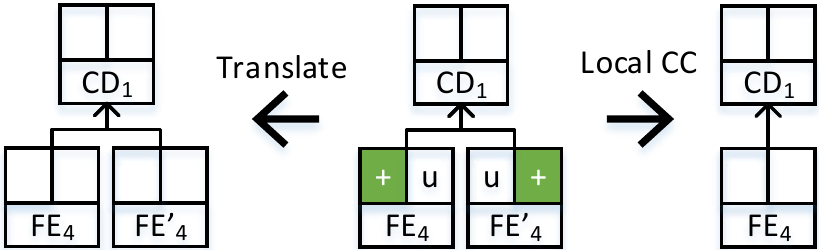}
	\caption{Running Example -- Translate vs. Local CC}
	\label{fig:rex_localCC}
\end{figure}

\paragraph{A worked out example of concurrent synchronization}
To correlate corresponding changes on both sides, we apply \emph{Local CC} first.
Nodes \emph{F$_4$} and \emph{E$_4$} are created independently with the same name but not set into correspondence yet.
Using \emph{Local CC}, we are also able to relate both changes to each other, creating the missing correspondence link in between. 
The effect on the precedence graph is shown on the right of \cref{fig:rex_localCC}.
On the left of this figure, we see the result of not using \emph{Local CC} but \emph{Translate}. 
Applying \emph{Translate} on both \emph{F$_4$} and \emph{E$_4$} independently of each other would create corresponding elements on the opposite sides using the proper forward and backward rule.
This step results in new elements \emph{F$_4^{\prime}$} and \emph{E$_4^{\prime}$} besides the former ones.

Next, we apply \emph{Translate} to several changes that are not contained in a conflict (scope) (as described in \cref{chapter:conflict_detection}).
This is the case for \emph{CD$_3$}, \emph{GL$_{14}$}, and \emph{GL$_{15}$}.
Translating newly added elements establishes consistency of each precedence node since the underlying rule applications are now complete and thus consistent.

\begin{figure}[h]
	\centering
	\includegraphics[width=0.7\linewidth]{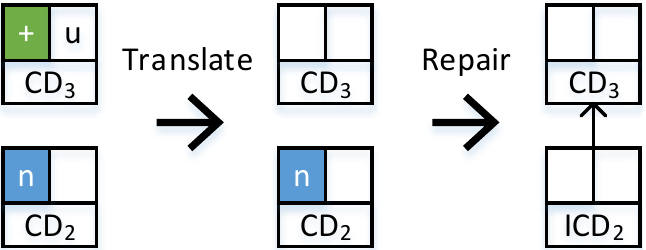}
	\caption{Running Example -- Translation then Repair}
	\label{fig:rex_repair_CD}
\end{figure}

Afterwards, we apply \emph{Repair} to fix broken rule applications such as \emph{CD$_2$} that is inconsistent due to a NAC violation.
\Cref{fig:short_cut_rule} depicts the shortcut rule \emph{CD-To-ICD}$_{\mathit{FWD}}$ that can be applied here to transform \emph{CD$_2$} into \emph{ICD$_2$}. The effect is that \emph{D$_2$} is preserved and an edge between \emph{D$_3$} and \emph{D$_2$} is created. The corresponding effect on the precedence graph is shown in \cref{fig:rex_repair_CD}. 
Note that the prior propagation of \emph{CD$_3$} is necessary to create the context needed by \emph{CD-To-ICD}$_{\mathit{FWD}}$ to be applicable.

\begin{figure}[h]
	\centering
	\includegraphics[width=\linewidth]{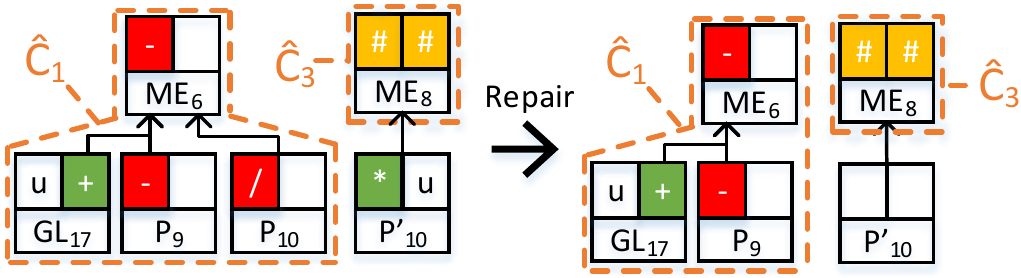}
	\caption{Running Example -- Reduce conflict scope $\hat{C}_1$}
	\label{fig:rex_repair_conflict_1}
\end{figure}

Now, only conflicts (scopes) remain that are resolved in any order using \emph{Resolve Conflict}.
The intermediate model and delta precedence graph can be found in \cref{chapter:conflict-res-ext}.
Starting with conflict scope $\hat{C}_1$, the primary issue is that a glossary link was created at entry  \emph{E$_6$} while deleting the method \emph{M$_6$} that corresponds to \emph{E$_6$}.
Considering the precedence node \emph{P$_{10}$}, only parts were deleted which implies that some of the remaining elements are to be preserved (here \emph{P$_{10}$}).
Using \emph{Repair} first allows us to reduce the size of $\hat{C}_1$ by propagating the re-location of \emph{P$_{10}$} from \emph{M$_6$} to \emph{M$_8$}. 
This step is depicted in \cref{fig:rex_repair_conflict_1}.

\begin{figure}[h]
	\centering
	\includegraphics[width=\linewidth]{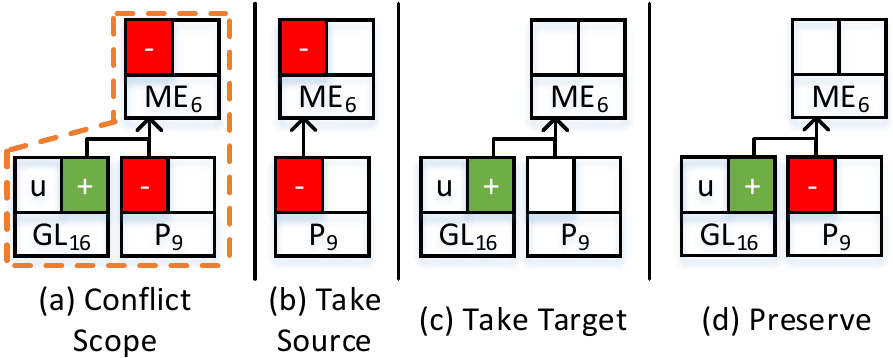}
	\caption{Running Example -- Resolving $\hat{C}_1$}
	\label{fig:rex_solving_conflict_1}
\end{figure}

Figures~\ref{fig:rex_solving_conflict_1} (b) -- (d) depict the results of all three conflict resolution strategies applied to conflict scope $\hat{C}_1$ in \cref{fig:rex_solving_conflict_1}~(a).
The application of \emph{Preserve} is of special interest as it revokes deletion deltas that block create deltas from being propagated.
Following that strategy, the changes to \emph{ME$_6$} are revoked to solve the conflict, while keeping the changes to \emph{P$_9$} untouched.
Applying any of these strategies leaves the remaining elements in a state where they can be propagated without colliding with any changes on the opposite side. 
To resolve $\hat{C}_1$ finally, we use \emph{Take Source} (as specified earlier) and revoke all changes on the target side that are related to the conflict.

\begin{figure}[h]
	\centering
	\includegraphics[width=\linewidth]{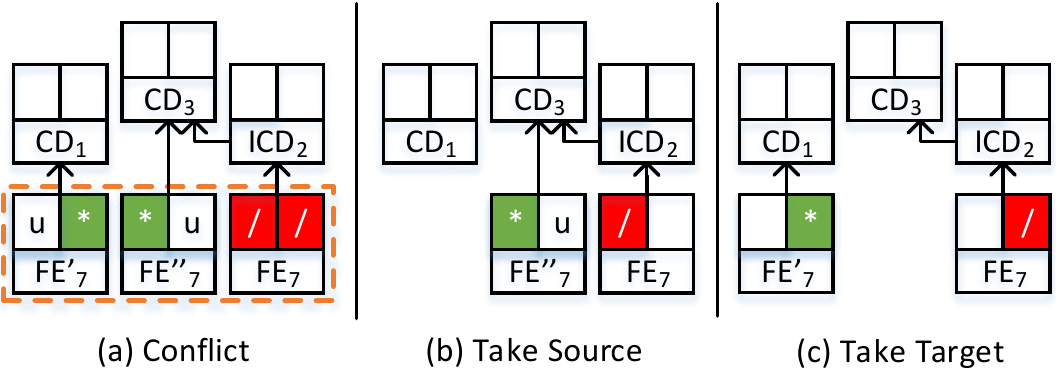}
	\caption{Running Example -- Resolving $\hat{C}_2$}
	\label{fig:rex_conflict2}
\end{figure}

For resolving the conflict scope $\hat{C}_2$, we perform \emph{Repair} first, but this application does not have any   effect here as both changes contradict each other. 
\Cref{fig:rex_conflict2} depicts the results of \emph{Take source} (b) and \emph{Take target} (c), which revoke the changes on one side.
Using the operationalized short-cut rule \emph{CD-To-ICD$_{\mathit{FWD}}$} after (b) or \emph{CD-To-ICD$_{\mathit{BWD}}$} after (c) propagates the changes to the opposite side, respectively.
Note that \emph{Preserve} has no effect here due to the nature of \emph{correspondence preservation} conflicts.
Since we chose \emph{Take Source} as conflict resolution strategy,  we can react to the relocation of F$_7$ to C$_3$ now by also moving E$_7$ from D$_2$ to D$_3$. This is done by applying \emph{Propagate} and thus applying \emph{CD-To-ICD$_{\mathit{FWD}}$}. 

Finally, we apply \emph{Repair} to conflict scope $\hat{C}_3$, which has no effect as both attribute changes contradict each other.
Resolving $\hat{C}_3$ can be done by choosing either \emph{Take source} or \emph{Take target}.
Again, \emph{Preserve} would not have any effect since there are no deletions that block the propagation of additions.
Instead, we have to decide which attribute change to propagate (which implies to revoke the opposite one).
Applying \emph{Take Source} and subsequently \emph{Propagate}, the remaining attribute change is propagated by re-evaluating the corresponding attribute constraint and transferring the value to E$_8$.

This leaves \emph{Clean up} with nothing to do since all changes have been accounted for.
The final model and its (delta) precedence graph (showing the result to be correct) are depicted in \cref{chapter:conflict-res-ext}. 

\section{Implementation}
\label{chapter:implementation}
\begin{figure}[h]
	\centering
	\includegraphics[width=\linewidth]{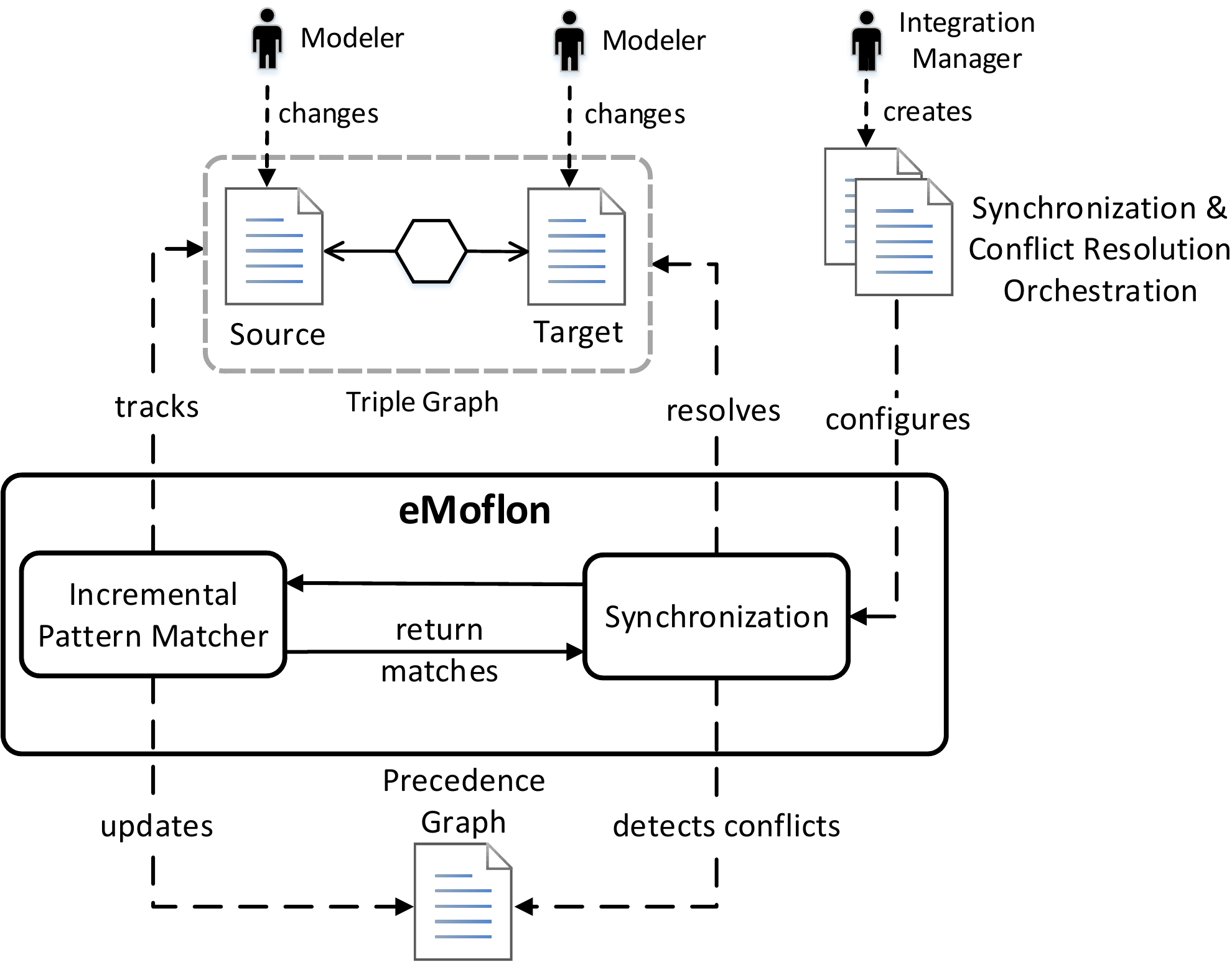}
	\caption{eMoflon - Synchronization Components}
	\label{fig:implementation}
\end{figure}

Our approach is implemented in a synchronization component as part of the state-of-the-art model transformation tool eMoflon~\cite{Weidmann19}. 
\Cref{fig:implementation} depicts the synchronization components with its interdependencies to an incremental pattern matcher and its inputs and outputs. 
In our synchronization framework, we allow modelers to change the source and target of a triple graph independently.
eMoflon keeps track of these changes by employing an incremental pattern matcher that throws events when new matches of TGG rules have been detected or existing ones have been invalidated.
This information is used to update a (delta) precedence graph which represents the dependencies between TGG rule applications. 
The synchronization component analyzes the delta precedence graph to detect conflicts. 
They can be resolved by a synchronization and conflict resolution orchestration that has been implemented by an integration manager, who is an expert in both source and target domain. 
However, we offer pre-defined configurations such as the one from our running example in the previous chapter that can be extended. 
Applying this orchestration resolves the previously detected conflicts step-by-step and restores consistency of the triple graph. 
eMoflon already has an extensive test suite\footnote{\href{https://github.com/eMoflon/emoflon-ibex-tests}{https://github.com/eMoflon/emoflon-ibex-tests.}} with 344 tests for various TGG-based consistency restoration scenarios from which 25 constitute concurrent synchronization tests.
In the near future, we will extend this test suite especially w.r.t. to more concurrent synchronization scenarios.

\section{Evaluation}
\label{chapter:evaluation}
In this section, we present our evaluation results, which are based on our implementation in eMoflon and two different TGG projects.
The first TGG is our running example consisting of 8 TGG rules, while the second one is based on the example introduced in \cite{FKST19} and consists of 28 TGG rules that define consistency between MoDisco~\cite{Modisco14} and custom documentation models.
However, the second TGG does not only contain more rules but also more asymmetric rules, i.e., rules that are no simple 1-to-1 mapping between source and target.
As a test environment, we use a workstation with an AMD Ryzen Threadripper 2990WX 3GHz 32xCore using 128GB of RAM.
One of the main goals of our approach is to provide a scalable concurrent synchronization solution.
Therefore, we pose the following research questions: 
\textbf{RQ1:} {\em Does our synchronization approach scale with the size of the model or with the number of changes and conflicts?}
\textbf{RQ2:} {\em Does the performance of the conflict detection change if less changes lead to actual conflicts?}
\textbf{RQ3:} {\em How does the number of changes and conflicts affect the conflict resolution performance?}

To answer these questions, we investigate two scenarios for each TGG:
First, we generate a fixed number of 100 conflicts and increase the size of both, the source and the target model, which we measure in number of nodes.
Second, we choose a fixed model size of about 500\,000 nodes (in the source and target models together) and increase the number of concurrent changes.
To investigate the correlation between changes and actual conflicts, we distinguish between 4 sub-scenarios by choosing the changes in such a way that 25\,\%, 50\,\%, 75\,\%, and 100\,\% lead to an actual conflict, respectively.
For all scenarios, we plot the initialization time where the incremental pattern matcher collects all matches for the still consistent triple graph, i.e., computes the precedence graph.
Then, we apply a number of changes to both, the source and target model and measure the time to restore consistency.
Note that each conflict-inducing change is meant to induce one of the three conflict types previously presented.
For each data point, we measured 20 repetitions and took the average value over all runs.

\begin{figure}[th]
	\centering
	\includegraphics[width=\linewidth]{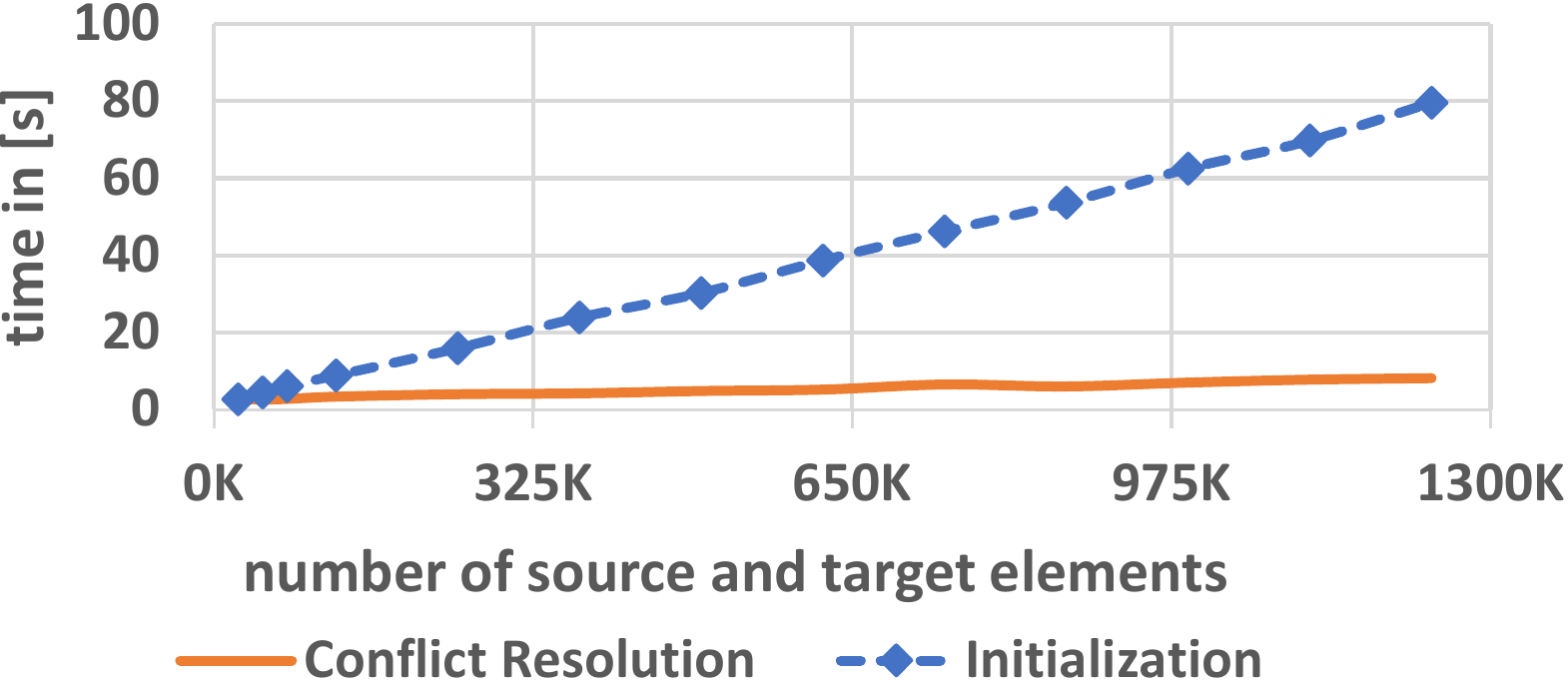}
	\caption{First TGG: Increasing model size with constant number of 100 conflicts}
	\label{fig:eval_const_conflict}
\end{figure}

\begin{figure}[th]
	\centering
	\includegraphics[width=\linewidth]{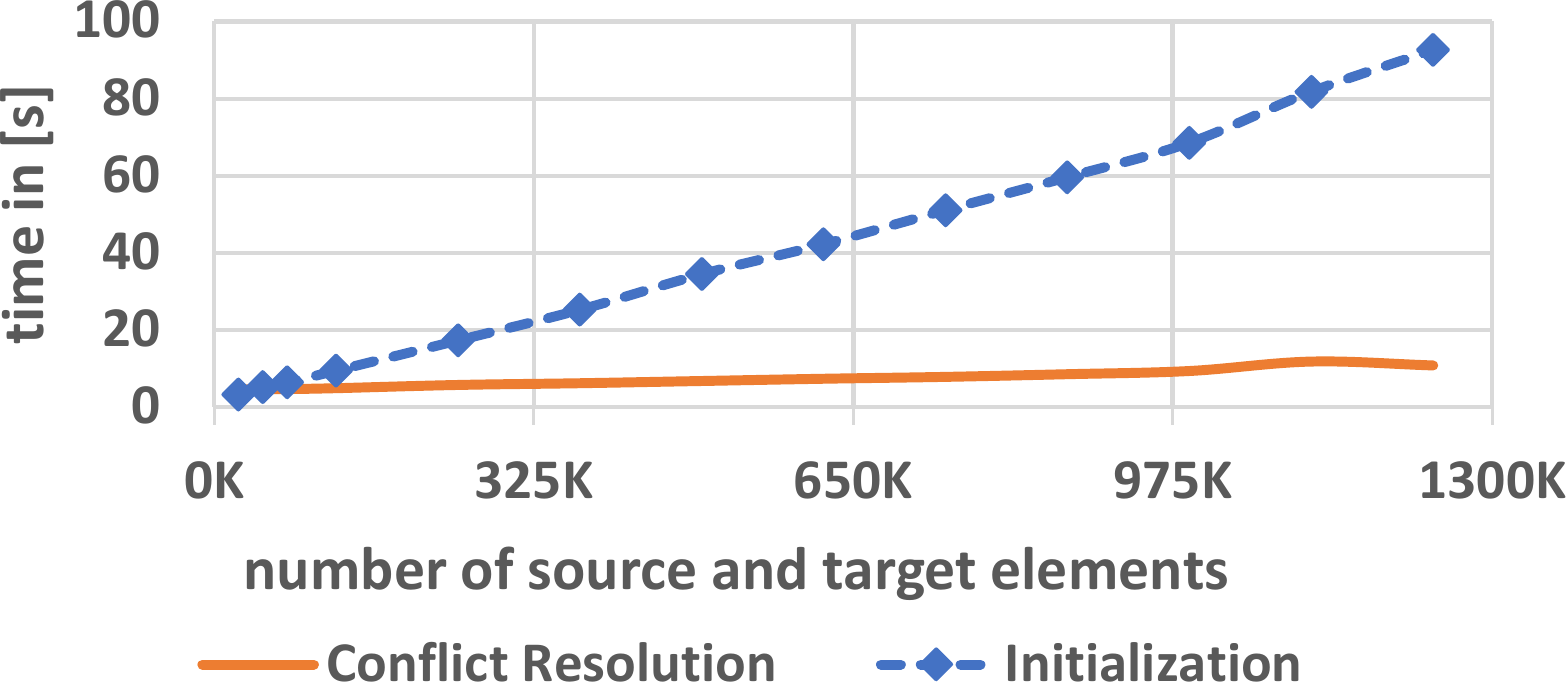}
	\caption{Second TGG: Increasing model size with constant number of 100 conflicts}
	\label{fig:eval_const_conflict_2}
\end{figure}

\begin{figure}[th]
	\centering
	\begin{tikzpicture}[spy using outlines={magnification=2, circle, size=1.1cm, blue, connect spies}]
	\node (n1) at (0,0) {\includegraphics[width=\linewidth]{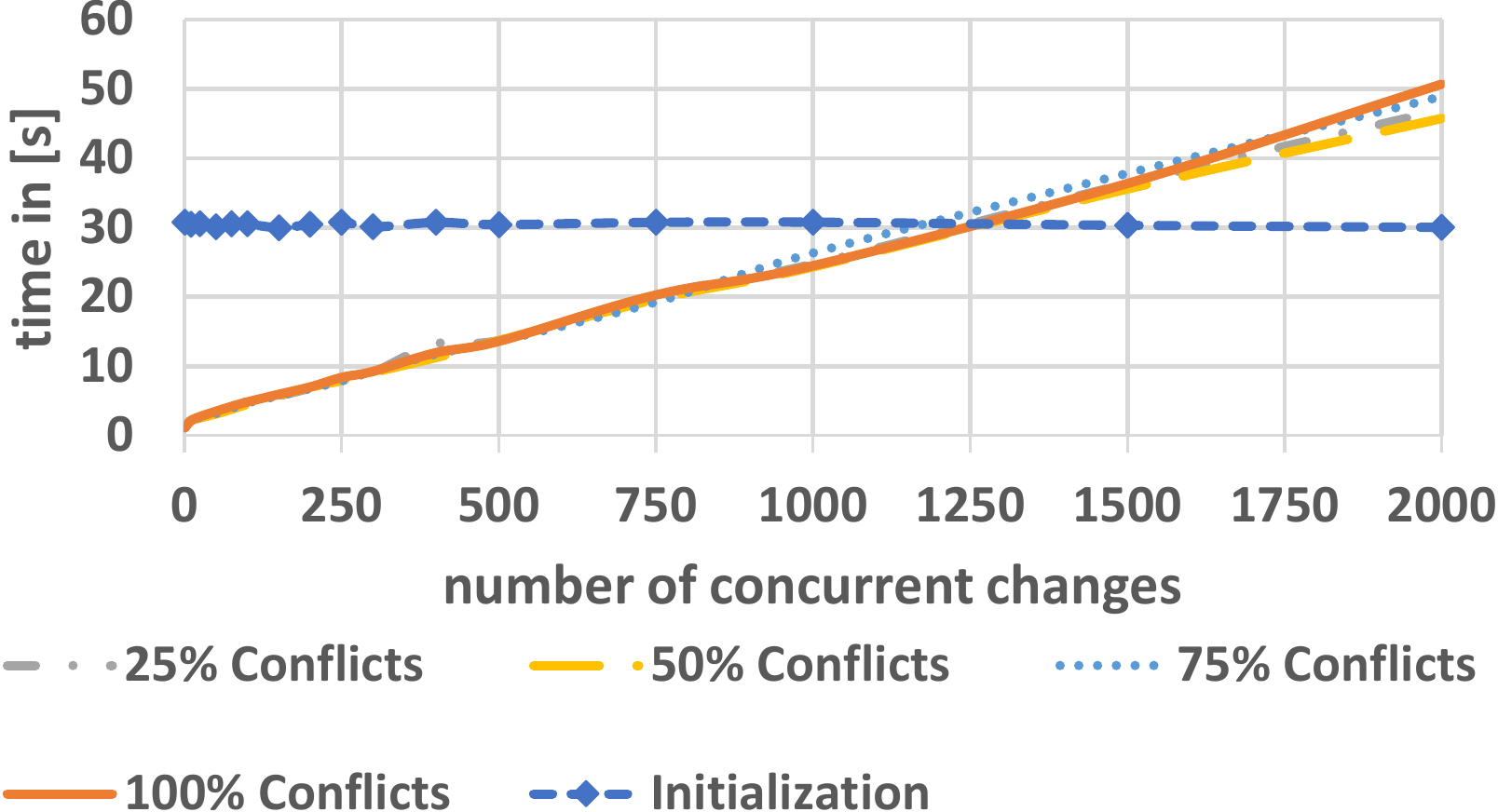}};
	\spy on (3.55, 1.58) in node at (2, 0.3);
	\end{tikzpicture}
	\caption{First TGG: Increasing number of changes with constant model of 500K nodes}
	\label{fig:eval_const_model}
\end{figure}

\begin{figure}[th]
	\centering
	\includegraphics[width=\linewidth]{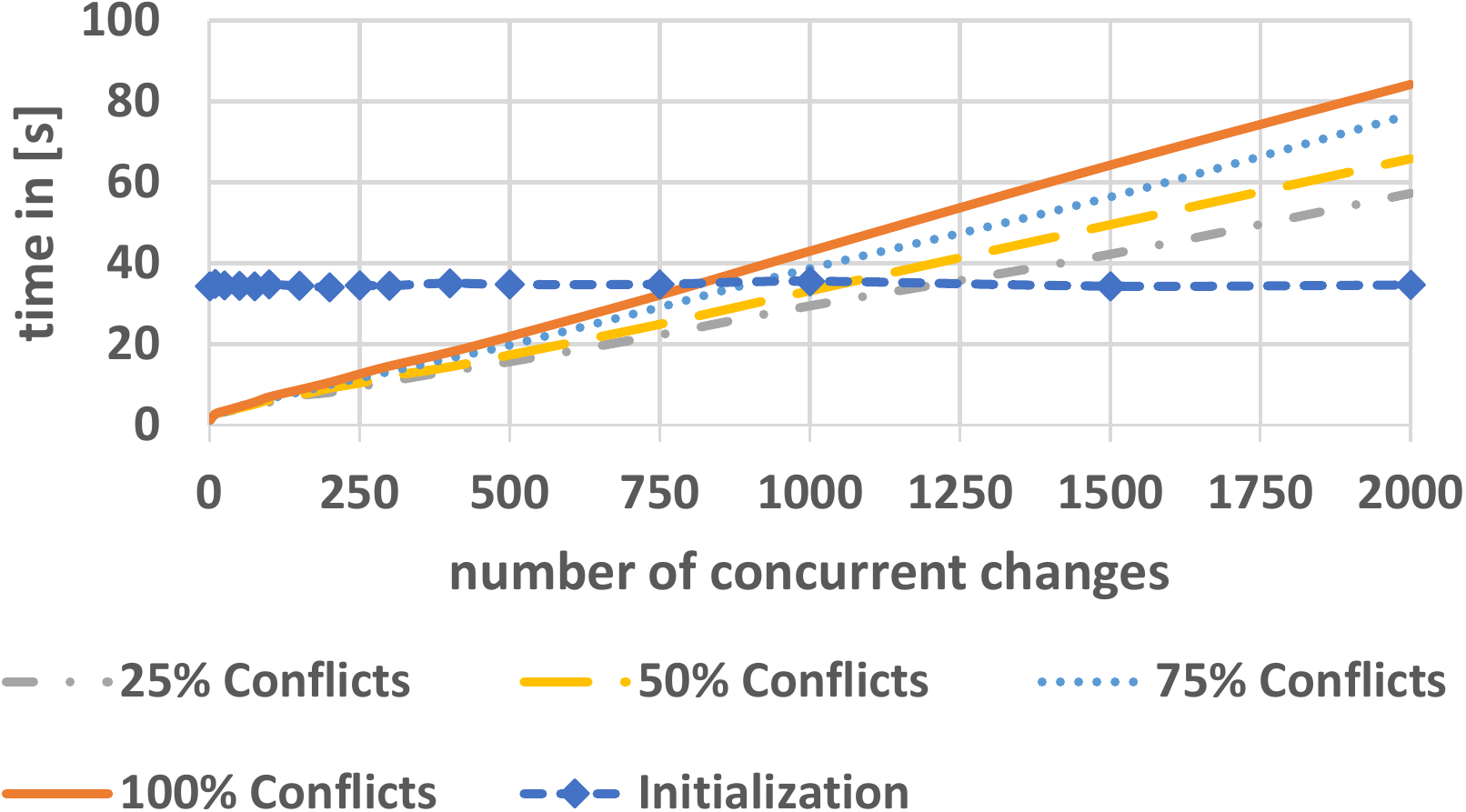}
	\caption{Second TGG: Increasing number of changes with constant model of 500K nodes}
	\label{fig:eval_const_model_2}
\end{figure}

\Cref{fig:eval_const_conflict} and \cref{fig:eval_const_conflict_2} depict the plots of the first scenario:
The initialization time linearly increases for the first TGG with the size of the model, while the conflict resolution time stays almost constant.\footnote{It slightly increases since deletions become more expensive with increasing model size in the Eclipse Modeling Framework.}
The same holds for the second TGG with a larger set of rules, which takes about 19\,\% longer to initialize.
This means that the model size does not directly affect the performance of conflict resolution \textbf{(RQ1)}.
\Cref{fig:eval_const_model} and \cref{fig:eval_const_model_2} depict the plots for the second scenario:
The initialization time stays constant with a constant model size of 500\,000 (source and target) nodes while the time to detect and resolve conflicts increases linearly.
Whether the changes to both sides are in conflict with each other has only a minor impact on the performance and increases the gradient slightly. 
For the larger set of rules, the impact becomes more significant and takes 16\,\% more seconds per 25\,\% more conflicts \textbf{(RQ2)}.
Thus, we can conclude that the performance scales linearly with the size of changes and to some point with the amount of conflicts that are introduced by changes \textbf{(RQ3)}. 
Note that the performance is only related to the chosen conflict resolution strategy in the way that it is more expensive to propagate many changes, e.g., not applying many deletions to one side in order to propagate one addition on the other would of course be less expensive than the other way around.

\paragraph{Threats to validity}
Our evaluation is based on synthesized models and changes only.
Thus, it remains future work to investigate real-world scenarios by analyzing, e.g., Git merge requests and deducing models from code.
Furthermore, we evaluated only with two TGGs that have different characteristics but describe a similar scenario.
However, the rules of both TGGs are symmetric as well as asymmetric and thus, do not only represent simple 1-to-1 mappings, which makes them representative for a broader range of TGGs.


\section{Conclusion}
\label{chapter:conclusion}
In this paper, we presented a scalable TGG-based, pre\-ce\-dence-driven concurrent model synchronization approach.
We \- showed how to use a delta precedence graph to identify conflicts and introduced a modular framework that enables developers to resolve these conflicts and, furthermore, orchestrate the whole process to achieve specific synchronization goals.
Giving a broad overview of the landscape of concurrent synchronization works, we showed that this approach is indeed novel in that we detect conflicts before propagating any changes.
Furthermore, we showed how to detect a new kind of conflict, namely \emph{correspondence preservation} conflicts, which to the best of our knowledge, no other approach is able to detect so far. 
More specifically, some approaches (e.g., ~\cite{Orejas20, MC16}) are also able to handle these situations, however, not in a self-reflective and transparent way as they only implicitly handle it.
Our approach has been implemented and evaluated within the state-of-the-art graph transformation tool eMoflon.
In the evaluation, we showed for two different TGG projects that our synchronization approach scales linearly with the size of changes instead of the model size.

For the future, we plan to extend and formalize our approach w.r.t. handling further types of conflicts between rule applications that, if applied, would violate, e.g., multiplicity constraints of a metamodel.
Further investigations are also needed to study the effects of one major design decision of our concurrent model synchronization approach in practice: still consistent rule applications remain untouched if they do not (causally) depend on a broken rule application. 
This design decision is inspired by \emph{a least change and least surprise principle}~\cite{CGMS17}  and is one main reason for the scaleability of our approach. 
Due to this, our synchronization algorithm is in general not always able to reestablish the consistency of two models when a change in one model requires its propagation against the introduced causal dependency relation in the related other model.
However, there are works such as \cite{KLKS10, Fritsche19} that show how to derive application conditions with a statically analyzable criterion for TGGs such that our synchronization algorithm very rarely runs into such a situation.
In our experience TGGs violate this criterion if and only if the propagation of a local change in one model would have rather unexpected and unwanted global effects on the other model from a user's point of view. 
But further experiments are needed to confirm our experiences.
Finally, we plan to build up a rich zoo of concurrent synchronization scenarios and to compare different approaches with respect to these. 

\begin{acks}
	We would like to thank the anonymous referees for their helpful suggestions. 
	This work is supported by the \grantsponsor{DFG}{German Research Foundation (DFG)}{https://www.dfg.de/en/index.jsp} under Grant No.:~\grantnum{DFG}{TA294/17-1}.
\end{acks}

\appendix


\section{Overview over Conflicts and Actions for Propagation}
\label{chapter:overview-conflicts-actions}
In \cref{tbl:overview-possible-conflicts} we give an overview of the different potential conflicts depending on the annotation of the DPG. 
Conflicts always concern nodes that stem from the original PG. 
The annotations in the DPG inform about possible kinds of conflicts such a node could be a part of. 
For example, a attribute change conflict can only occur when both source and target annotation of a node contain $\#$. 
In contrast, whenever a node is annotated with $/$ (no matter if at source, target, or both), it has to be checked whether this node participates in a preserve-delete or correspondence preservation conflict.

\begin{table*}
	\caption{Overview of potential conflicts depending on the annotations in the DPG where \emph{acc} stands for attribute change, \emph{pdc} for preserve-delete, and \emph{cpc} for correspondence preservation conflict.}
	\label{tbl:overview-possible-conflicts}
	\centering
	\begin{tabular}{lccccc}
		\toprule
		\diagbox{source}{target}	& $\{\}$		& $-$				& $/$				& $\#$			& n \\
		\midrule
		$\{\}$										& 					& pdc				& pdc, cpc	&						& cpc \\
		$-$												& pdc				&						& pdc, cpc	& pdc				& pdc, cpc \\
		$/$												& pdc, cpc	& pdc, cpc	& pdc, cpc	& pdc, cpc	& pdc, cpc \\
		$\#$											& 					& pdc				& pdc, cpc	&	acc				& cpc\\				
		n													& cpc				& pdc, cpc	& pdc, cpc	& cpc				& cpc\\
		\bottomrule
	\end{tabular}
\end{table*}

\section{Intermediate Conflict Resolution Steps}
\label{chapter:conflict-res-ext}
\begin{figure*}[th]
  \centering
  \includegraphics[width=\linewidth]{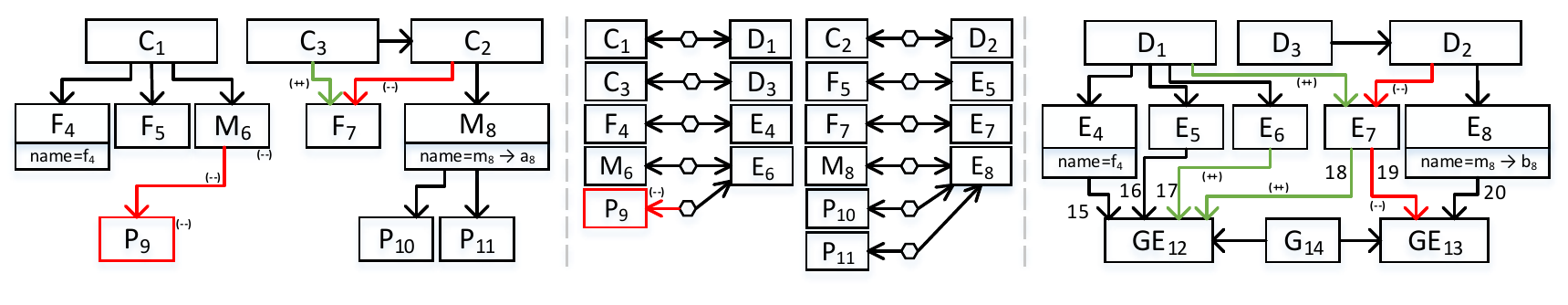}
  \caption{Running Example -- Model and Delta after LocalCC, Repair and Translate}
  \label{fig:model_before_conf_res}
\end{figure*}

\begin{figure*}[h]
	\centering
	\includegraphics[width=.8\linewidth]{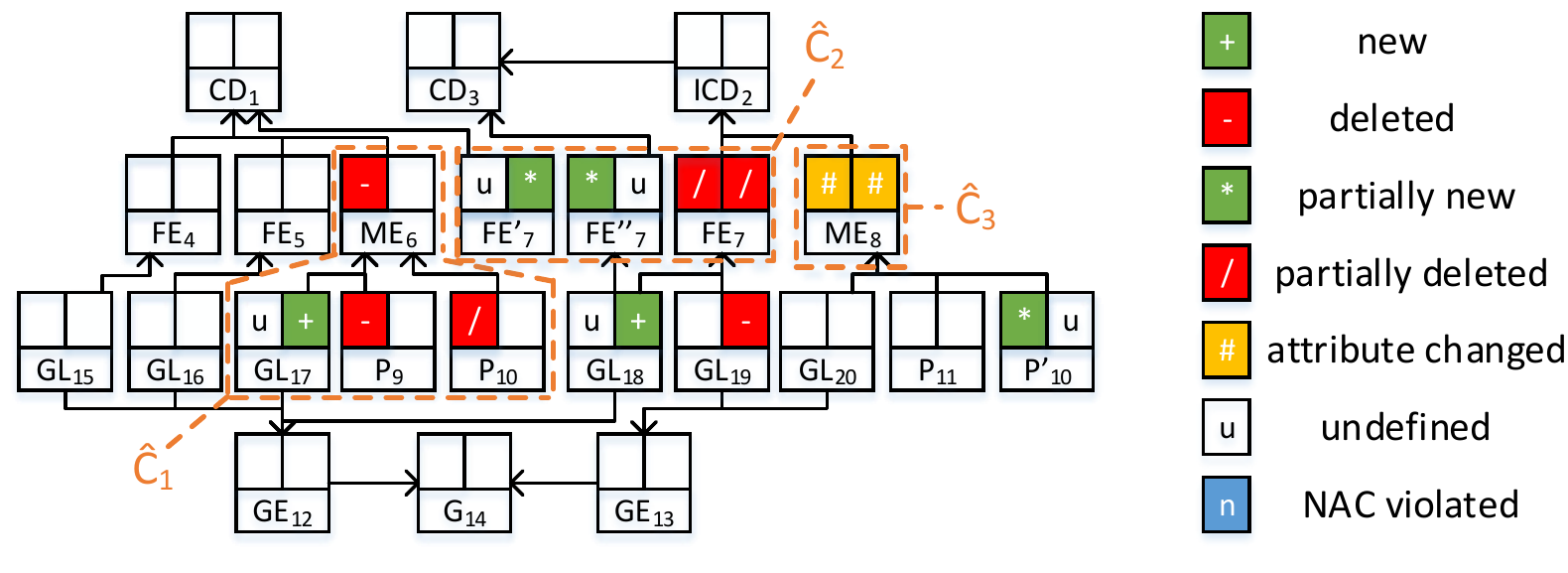}
	\caption{Running Example -- Delta Precedence Graph after LocalCC, Repair and Translate}jo
	\label{fig:pg_before_conf_res}
\end{figure*}

\begin{figure*}[th]
	\centering
	\includegraphics[width=\linewidth]{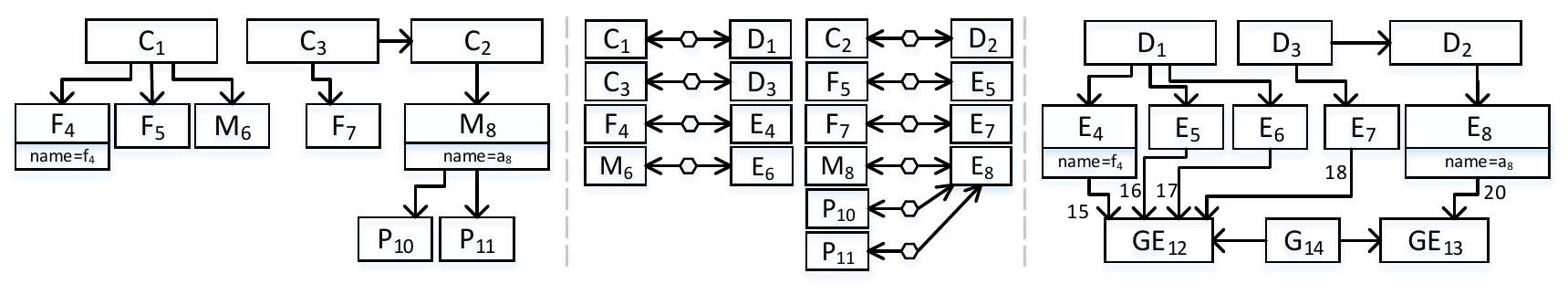}
	\caption{Running Example -- Synchronized Model}
	\label{fig:model_result}
\end{figure*}

\begin{figure*}[h]
	\centering
	\includegraphics[width=.8\linewidth]{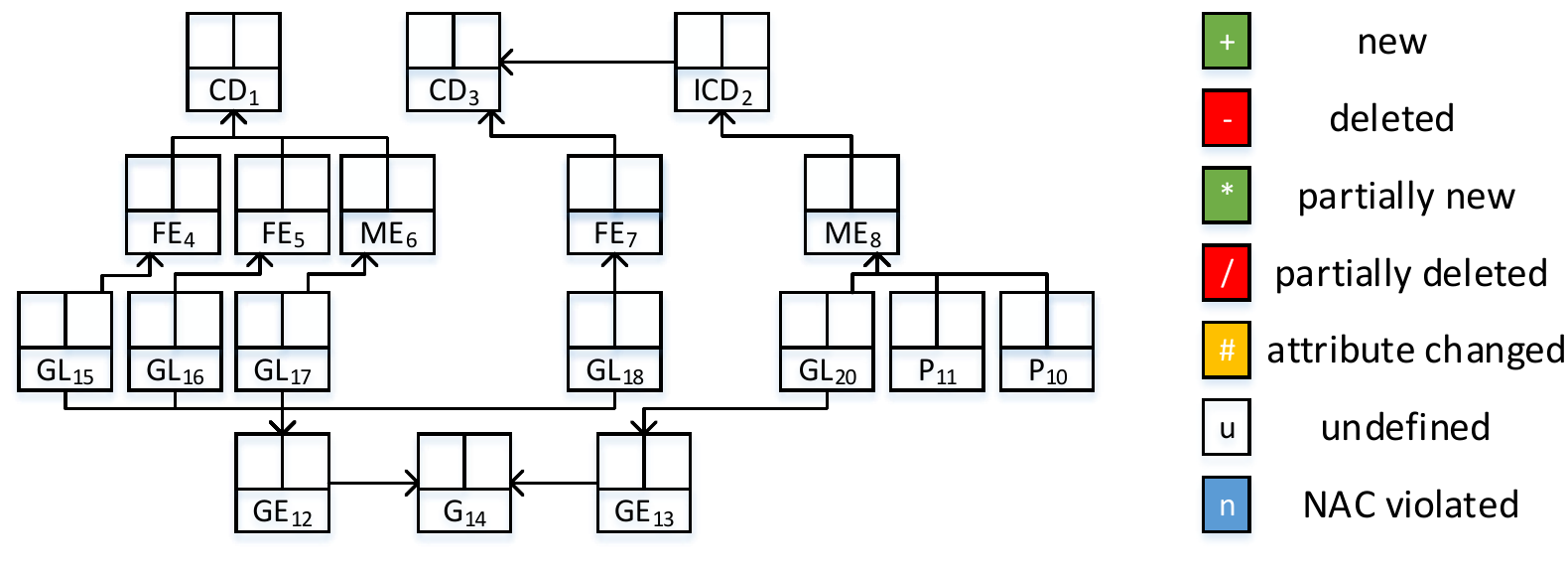}
	\caption{Running Example -- Synchronized Precedence Graph}
	\label{fig:pg_result}
\end{figure*}

\Cref{fig:model_before_conf_res} depicts the model of our running example after \emph{Local CC}, \emph{Translate} and \emph{Repair} have been applied, while \cref{fig:pg_before_conf_res} shows the corresponding delta precedence graph with conflicts $\hat{C}_1$, $\hat{C}_2$ and $\hat{C}_3$.
\Cref{fig:model_result,fig:pg_result} depict the final model and delta precedence graph, respectively.

\clearpage



\bibliographystyle{ACM-Reference-Format}

\end{document}